\documentclass[useAMS,usenatbib,usegraphicx,usedcolumn,authoryear]{elsarticle}
\usepackage{url}
\usepackage{amsmath}

\begin{document}
\begin{frontmatter}

%
%
%


\title{On the prospects of Near Earth Asteroid orbit triangulation using the Gaia satellite and Earth-based observations}


\author[sigi]{Siegfried Eggl}  
\author[enrico]{Hadrien Devillepoix}

\address[]{IMCCE Observatroire de Paris, UPMC, Universit\'e Lille 1, 
				  75014 Paris, (France)}

\end{frontmatter}
				 
\begin{abstract}
Accurate measurements of osculating orbital elements are essential in order to understand and model the complex dynamic behavior of Near Earth Asteroids (NEAs).
ESA's Gaia mission promises to have great potential in this respect. In this article we investigate the prospects of constraining orbits of newly discovered and known NEAs
using nearly simultaneous observations from the Earth and Gaia.  
We find that observations performed simultaneously from two sites can effectively
constrain preliminary orbits derived via statistical ranging. 
By linking discoveries stored in the Minor Planet Center databases to Gaia astrometric alerts one can identify nearly simultaneous observations of
Near Earth Objects and benefit from improved initial orbit solutions at no additional observational cost.
\end{abstract}


%


\section{Introduction}

Both, the airburst of a bolide over Chelyabinsk/Russia on Feb. 15$^{th}$, 2013  and the deep close encounter of the asteroid 2012 DA14 missing the Earth by as little as 3.5 Earth radii  \cite{jpl-sentry-2012} 
 highlighted once more that Near Earth Object (NEO) pose a non-negligible threat to mankind. 
Predicting future encounters between asteroids and the Earth should, therefore, be considered a task of high priority. 
Yet, current estimates project that only about 30\%  of the total NEO population with diameters between $100$m and $1$km have been discovered so far \cite{mainzer-et-al-2012}.
This issue is further aggravated by the fact that discovering an asteroid does not automatically entail knowledge on whether it will collide with the Earth or not.
Due to the complex interplay of gravitational and non-gravitational forces, long term predictions of NEA impact risks are a difficult task, especially
when initial orbits\footnote{the osculating orbital elements that are derived when a NEO is first discovered} are poorly constrained.
Regarding discovery and orbit improvement, the Gaia astrometry mission \cite{mignard-et-al-2007} has been found to hold great potential for NEA research
\cite{bancelin-et-al-2010,hestroffer-et-al-2010,tanga-mignard-2012}.
Given the fact that Gaia is a space observatory with a fixed scanning law, however, consecutive
observations of newly discovered objects, which are vital for initial orbit determination, are necessarily sparse. In practice this means that many objects have to be followed up from ground based sites \cite{thuillot-2011}.
Since only directional data are available from astrometric observations, preliminary orbital elements and ephemeris are generated using statistical ranging (SR) \cite{virtanen-et-al-2001}.
The resulting orbits are used to project a NEO's future position on the sky plane to facilitate follow up observations. 
Due to down-link schedules and data processing, the delay between a discovery alert and a follow up can be as large as 48 hours.
If the uncertainties in the initial orbits are large, the range of possible locations for the target NEO tends to grow very quickly.
Retrieving newly discovered objects might become difficult in such cases.
In this work we assess the potential of nearly simultaneous\footnote{Observations do not have to be exactly simultaneous. Asynchronicities that do not lead to a discrepancy in the 
observed astrometric position greater than the instrumental precision are permissible.} NEO observations from Gaia and the Earth to tackle this issue.
A combination of Gaia data and ground based observations has already been found to greatly enhance the quality of orbit predictions  \cite{bancelin-et-al-2011}. Yet, 
simultaneous TR of one object from two different locations was not considered in this work.
In contrast, several authors have proposed asteroid orbit improvement based on multiple observing stations considering a combination of space-based and Earth-bound observation sites \cite{gromaczkiewicz-2006, granvik-et-al-2007,chubey-et-al-2010, eggl-2011}.
Exploring techniques to link observations from sites with a considerable parallax, \cite{granvik-et-al-2007} gave hints that 
simultaneous observations can be favorable for constraining initial orbits.
\cite{gromaczkiewicz-2006, chubey-et-al-2010, eggl-2011} showed that independent orbit determination for asteroids via basic trigonometry is possible if
simultaneous observations from two sites are available.
%
%
%
%
%
%
We shall discuss some of the benefits and issues of simultaneous observation from Gaia and the Earth in the next sections.

%
%
%

  
  \section{Identification and Linking}
  \label{eggl:sec:idl}
\begin{figure}[t]
\centering\includegraphics[scale=0.5]{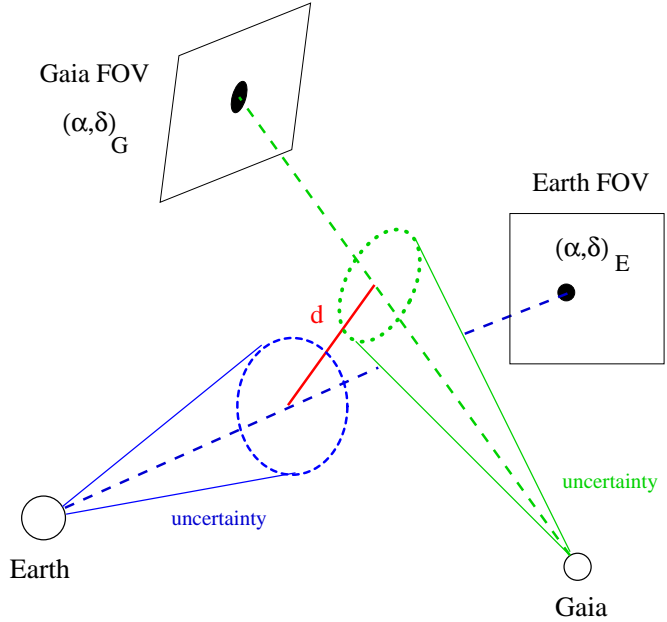}
\caption{The triangulation setup allows to link objects in both observer frames without the need to construct and compare individual sets of orbital elements per site, if the observations are 
nearly synchronous. Astrometric right ascension ($\alpha$) and declination ($\delta$)
pairs corresponding to the target's position in the respective FOVs can be used to construct geometric lines that have the observing stations at their origin.
If the distance (d) between those lines is smaller than the combined radius of the astrometric uncertainty ellipses at the point of closest approach,
objects recorded by the two observers can represent the same NEO.
\label{eggl:fig1}}
\end{figure}

Let us assume that nearly simultaneous observations from Gaia and at least one additional ground-based site have been performed.
Given the substantial parallax between the two observers, the astrometric FOVs can differ significantly. 
Hence, determining whether an object observed from both sites is in fact the same asteroid becomes crucial. 
For cataloged objects ephemeris predictions are mostly accurate enough for this purpose. 
The correct linking of observations of newly discovered objects can be more difficult. 
Should a sufficient number of observations be available, orbital element bundles can be constructed for each observation site. 
A comparison of the orbital element probability density functions generated
via orbital inversion of single night sets can then be used to identify and link the same objects in each frame. 
This so-called ephemeris-space multiple-address-comparison (eMAC) method has been suggested by \cite{granvik-muinonen-2005}. 
\cite{granvik-et-al-2007} showed that this technique works for observations with large parallaxes.
If observations are too sparse to generate orbital element bundles for each site individually, one can try to find observations from both sites that are nearly simultaneous.
 In this case one can assume that each site has a pair of right ascension ($\alpha$) and declination ($\delta$) values
 for each object recorded at approximately the same time. Timing errors that lead
 to deviations smaller than the astrometric precision of the observing instrument are acceptable. These ($\alpha$,$\delta$) pairs yield directional unit vectors that point 
 from the respective observer to the NEO. In the absence of strong gravitational fields, straight lines can be constructed
 from such astrometric data, connecting the observation sites to the target, see Figure \ref{eggl:fig1}.
 The distance $d$ between those lines is given by 
 \begin{align}
d=|(\vec{e}_E \wedge \vec{e}_G)\cdot({\vec{r}_G}-\vec{r}_E)|
\end{align}
where $\vec{r}_E$ and $\vec{r}_G$ represent the heliocentric position of the Earth and Gaia respectively, and
$\vec{e}_E$ and $\vec{e}_G$ are the corresponding line unit vectors
\begin{align}
\vec{e}_{E,G} =(\cos \alpha \cos \delta, \sin \alpha \cos \delta, \sin \delta)^T_{E,G}.
\end{align}
 If the distance between the two lines is smaller than the sum of the radii of the astrometric uncertainty ellipses evaluated at the point of closest approach,
 the two observations can be attributed to the same object.
 Of course, this is only a necessary, not a sufficient linking condition.

\section{Initial Orbit Determination}
  The possibility to constrain the location of an asteroid at a given epoch constitutes the main advantage of synchronously recorded parallactic data sets.
  As discussed in the previous section, this can be implicitly achieved by performing a classical or SR based orbit determination for both sites individually. 
  The resulting orbital element distributions can be cross-matched and outlying solutions excluded.
  This, however, requires a sufficient number of observations.
  Since no delay due to Gaia's alert time has to be taken into account in such a scenario, simple trigonometry can be used to confine possible NEO locations for the time of observation.  
  The advantage of TR compared to combining
  multiple 'one-site' observations lies in the fact that orbital constraints can be constructed 
  with as little as one synchronously recorded frame per site (see section \ref{eggl:sec:statr}).
  Having an Earth-based telescope observe 
  the same field of view as Gaia at specific times\footnote{e.g. whenever the FOV approaches the ecliptic} would, thus, allow for an improvement of initial orbits of newly discovered asteroids.
  It is questionable, however, whether telescope time would be made available for a program that stares along Gaia's line of sight without a predefined target. 
  Fortunately, active sky surveys such as Pan-STARRS will produce data sets that are temporally and spatially overlapping with Gaia observations. 
  As these observations are available via the Minor Planet Center, they can be combined with Gaia astrometry to constrain
  possible orbit solutions. 
  Even an independent epochal orbit determination via TR is possible, if the object is recorded in more than one consecutive frame by both observation sites \cite{eggl-2011}.
  In other words, initial orbital elements can in principle be created with only two frames per observing station. 
  
\section{Orbit Refinement}  
  Many initial NEO orbits had to be constructed based on very few observational data points. Due to the gravitationally active near Earth environment, positioning uncertainties tend to grow rather quickly for NEOs,
  and followup observations become essential in order not to lose track. Of course, once the Gaia mission has been completed and all observations
  have been reduced with the new catalog, the quality of orbits derived via Gaia data will be substantially better than anything achievable from ground based observations \cite{bancelin-et-al-2011}.
  However, up to that point, simultaneous observations can still be useful to improve NEO orbital elements.
  Having an initial orbit estimate at our disposal, it is possible to predict future Gaia-FOV crossings of NEOs.  
  Given Gaia's relatively large FOV, even asteroids on orbits with large uncertainties should be recoverable. Earth-based observations could then be conducted
  simultaneously with Gaia-FOV crossings, so that positioning of the asteroid via TR becomes possible. 
  Strictly speaking, TR is not necessary in this case, since the additional observational data from the second site does itself contain the constraints on the NEO's
  orbit. However, we will use TR instead of a statistical ranging or differential corrections technique in order to study the impact of 
  a readily accessible ranging and localization of the target in coordinate space on orbital elements.

\section{NEA Positioning Via Triangulation}
\label{eggl:sec:pos}
\begin{figure}[t]
\centering\includegraphics[scale=0.3]{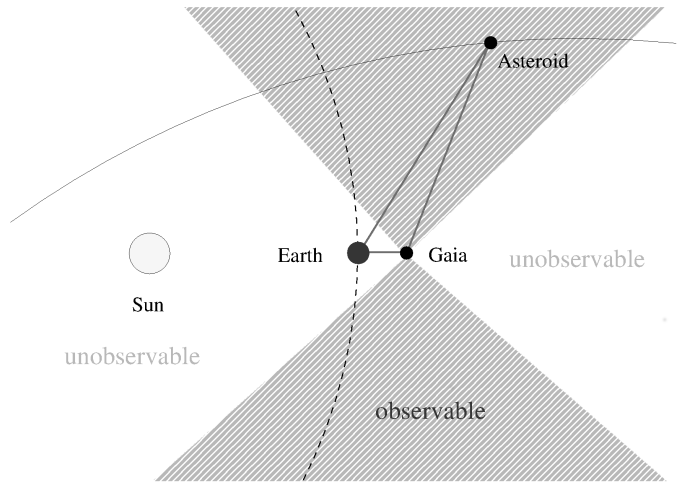}
\caption{Schematic of the triangulation setup projected into the Earth's orbital plane. 
All non-shaded regions are inaccessible to the triangulation process. We assume that all regions that can be observed by the Gaia satellite
are simultaneously accessible from the Earth.\label{eggl:fig1b}}
\end{figure}

Triangulation (TR) from observatories based on different sites is a fairly common tool in meteroid orbit determination \cite{ceplecha-1987}. 
The application of TR to refine NEA orbital elements has been investigated by \cite{gromaczkiewicz-2006} for two satellites in the Lagrangian points $L_4$ and $L_5$ of the Sun-Earth system. 
The approach was extended in \cite{eggl-2011} to encompass free satellite positioning and Earth based observations. Their results suggest that given two observations with sufficient spatial as well as temporal resolution, 
accurate orbital elements can be derived without having to rely on previous data or orbit-fitting models. 
We will apply the method proposed in \cite{eggl-2011} to evaluate the potential merits of NEA TR given Gaia and Earth-based observations.
Let us assume we have two sets of pairwise orthogonal angles, e.g. $(\alpha_G,\delta_G), (\alpha_E,\delta_E)$ - where the subscripts stand for Gaia and the Earth respectively - and the distance between the observing locations $d_{GE}$, all measured at the same instant. 
It is then possible to reconstruct the position of an observed object in a locally Euclidean frame of reference (EFOR). Let us furthermore choose the EFOR to be heliocentric. 
The fact, that the observed object has to be accessible from both 
sites - Gaia and the Earth - at the same time is one of the most limiting factors for TR. 
As Gaia's scanning 
law prohibits observations in two $45^\circ$ cones centered at $L_2$ (Sun-Earth) with axes towards and away from the Sun \cite{hestroffer-et-al-2010}, 
the accessible region for TR is rather restricted, see Figure \ref{eggl:fig1b}.

%

\begin{figure}[t]
\centering\includegraphics[scale=1]{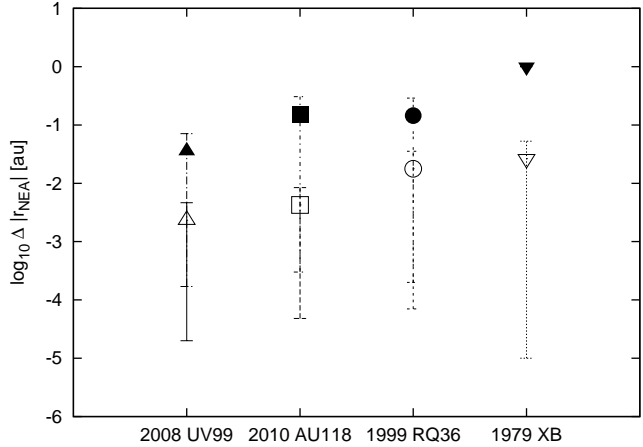}
\caption{Median and median deviation (error bars) values for 3D positioning errors of 4 NEAs using triangulation with Gaia and Earth-based observations.
The empty symbols denote best case results and the full symbols depict simulated results for worst-case observational uncertainties, see Table~\ref{eggl:tab2}.  \label{eggl:fig2}}
\end{figure}

Nevertheless, due to the configuration's proper motion, most of the Earth orbit crossing NEA population brighter than H magnitude $20$ should be observable during the 5 year 
mission.\footnote{A potential shift of the Gaia mission's cutoff H magnitude to $21$ is currently debated. Going to fainter magnitudes would make more NEOs accessible to triangulation assuming that most of the ground based 
sites can also observe the fainter objects with reasonable astrometric precision. We will continue assuming a conservative magnitude limit of $H=20$mag, however.}
In order to maximize the uniformity of sky coverage Gaia will perform 6-hour great circle scans, 
where the precessing spin axis retains a $45^\circ$ angle to the Sun at all times. A full precession cycle requires 63 days (see e.g. \cite{hestroffer-et-al-2010}).
Having a fixed scanning law, Gaia's $0.7^\circ \times 0.7^\circ$ field of view (FOV) cannot be altered to accommodate special observation schedules. 
Hence calculating rendezvous times between Near Earth Objects and Gaia's FOV becomes necessary. This is not an easy task, however, 
because it requires precise knowledge on the thermal and bulk properties of the satellite as well
as the exact initial positioning and attitude at $L_2$ after launch. Even-though tools like DPSC's rendezvous simulator do exist allowing to model FOV crossings, 
the scanning law's initial phase remains unknown.  
In the following we will, therefore, resort to a simplified statistical approach to determine the quality of positioning data achievable with TR.
In order to investigate the quality of NEA positioning via TR, 
 observations of four NEAs chosen form NASA JPL's Sentry risk table \cite{jpl-sentry-2012} are modeled, see Table~\ref{eggl:tab1}
The aim is to estimate the median quality of positioning of individual asteroids achievable during the 5 year Gaia mission.  
Using an $8^{th}$ order symplectic integrator \cite{yoshida-1993} the NEA orbits were propagated together with the 8 planets of the Solar System,
the Pluto-Charon center of mass, as well as  Ceres, Vesta, and Pallas. The Gaia satellite was positioned and kept exactly at $L_2$ (Earth-Sun), and no relativistic or non-gravitational forces were considered.
All initial osculating elements were taken from JPL-HORIZONS system for Jan $1^{st}$, 2014. The total integration time was set to 5 years corresponding to Gaia's planned mission duration.  
As the actual dates of NEA FOV crossings are dependent on the initial phase of Gaia's scanning law, the following statistical approach has been chosen:
A measurement process was simulated every $\Delta t=106.5$ minutes\footnote{See section \ref{eggl:sec:orbit} for details.}.
Measurements were accepted when the respective asteroid was in the accessible region for TR, and it had an apparent magnitude smaller than $v=20$mag.  
Here, we assumed that the dominating uncertainties are due to:
\begin{itemize}
\item the finite angular resolution in determining $(\alpha_G,\delta_G), (\alpha_E,\delta_E)$,
\item the uncertainty in the momentary distance between Earth and the Gaia satellite,
\item and the observed positioning offset of the asteroid due to finite light speed (see light travel time offset in \cite{eggl-2011}) which depends on the asteroid's velocity and the observers' positions relative to the asteroid.  
\end{itemize}  
Furthermore it is assumed, that the Earth's as well as Gaia's velocities in the EFOR are well known, so that differential and absolute aberration effects can be corrected for and
do not have to be considered in this setup. Gaia's attitude noise ($10^{-8} - 10^{-7}rad$) \cite{keil-theil-2010} is also assumed to be correctable a posteriori using relative positioning with respect to the stellar background. 
Simultaneous Earth-based measurements are assumed to be able to achieve the same limits in apparent magnitude as Gaia.
To assess the apparent magnitude of the asteroids investigated, their absolute magnitudes were extracted from \cite{jpl-sentry-2012}. 
A model of ideal diffusely reflective spheres has been adopted to evaluate the reflected light at any given phase angle following \cite{bowell-et-al-1989}.
Asteroids in the TR setup's accessible region are, of course, not automatically within Gaia's FOV. Therefore, the Ansatz is to estimate 
the median accuracy an observation would produce, if the asteroid passed the FOV once during the 5 year mission taking the dominating uncertainties into account. 
In Table~\ref{eggl:tab2} uncertainty values for a best-case as well as a worst-case scenario are provided. Gaia's performance values 
were extracted from the mission's data sheets\footnote{Specific data regarding the Gaia mission were acquired from \url{http://rssd.esa.int/gaia}, 2012.} as well as from \cite{hestroffer-et-al-2009}. 
Observation residuals listed in the Minor Planet Center database\footnote{\url{http://www.minorplanetcenter.net/iau/special/residuals.txt}, 2013},
suggest Earth-based astrometric precision to 
range from roughly $0.1"$ to around $2"$. 
The simulated TR results are compared to the NEAs' actual positions calculated via numerical orbit integration. The median and median deviation of the 
relative error in positioning is given in Figure \ref{eggl:fig2}. As expected, the best and worst case scenarios are separated by an order of magnitude, which is mainly due to 
the change in the quality of Earth-based observations. The quality of positioning is non uniform for the four NEAs considered. Even-though the results for 
2008 UV99 are quite promising, Earth-based observations with a precision of $2"$ make a TR of 1979XB practically impossible. 
Yet, in most cases positioning uncertainties between 0.001 and 0.1 au seem achievable. The implications for NEA orbit determination and refinement
will be discussed in the next two sections.

\section{Refinement of NEA Orbital Elements}
\label{eggl:sec:orbit}
\begin{figure}[t]
\begin{tabular}{cc}
\includegraphics[scale=0.66]{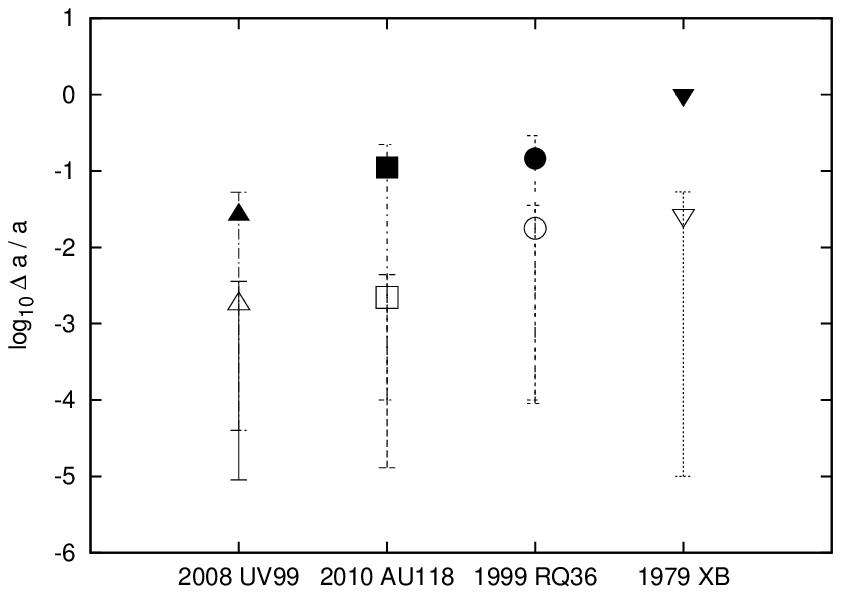} & \includegraphics[scale=0.66]{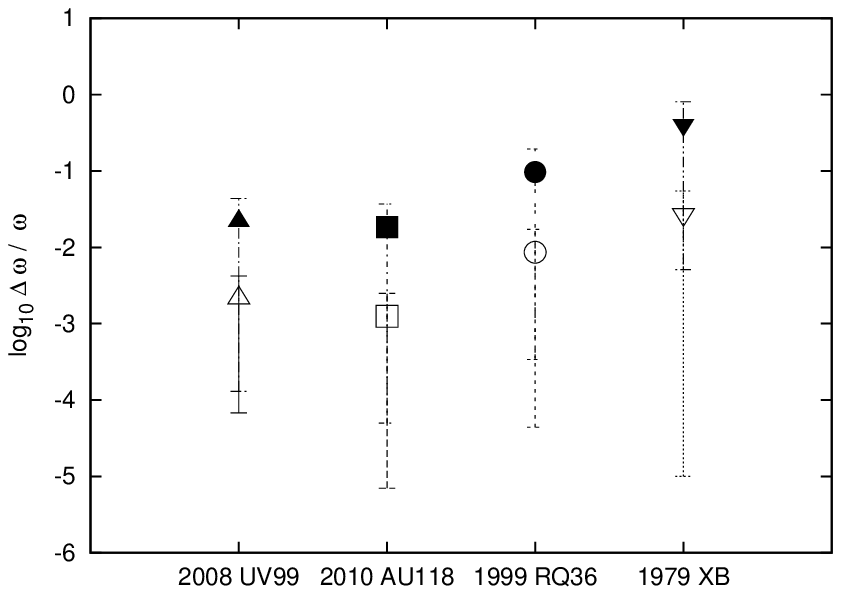} \\
\includegraphics[scale=0.66]{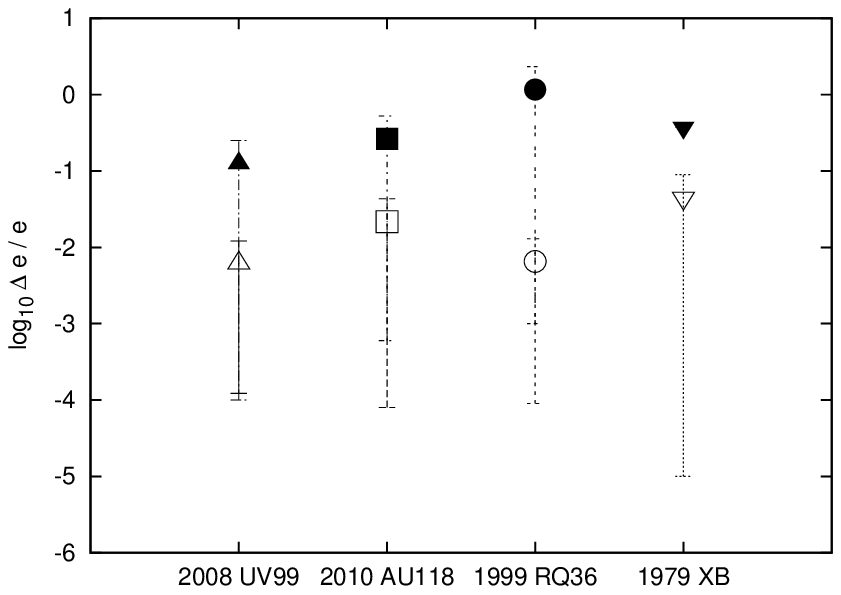} & \includegraphics[scale=0.66]{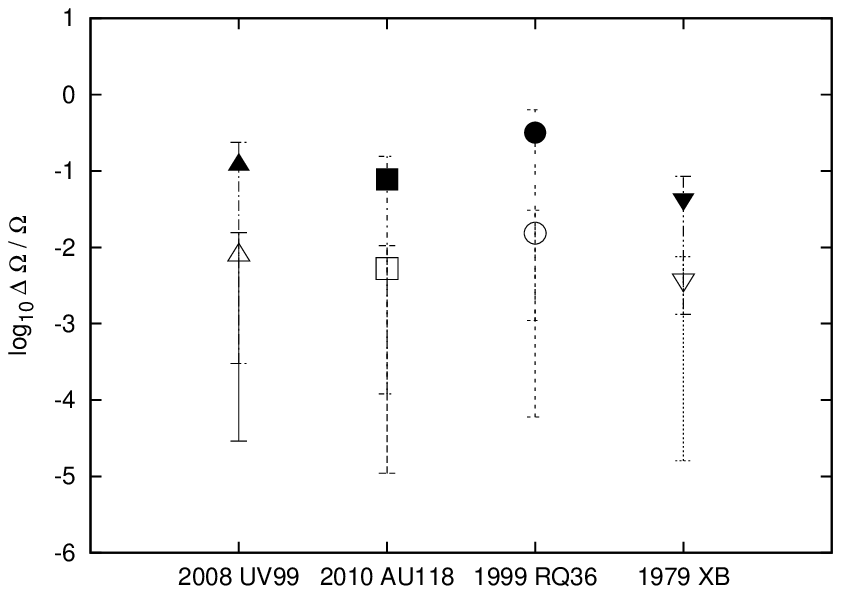} \\
\includegraphics[scale=0.66]{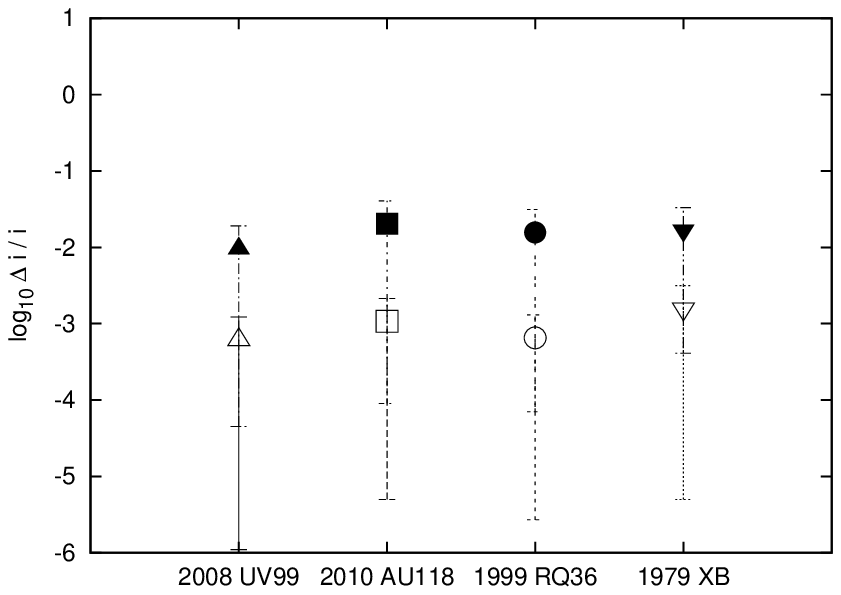} & \includegraphics[scale=0.66]{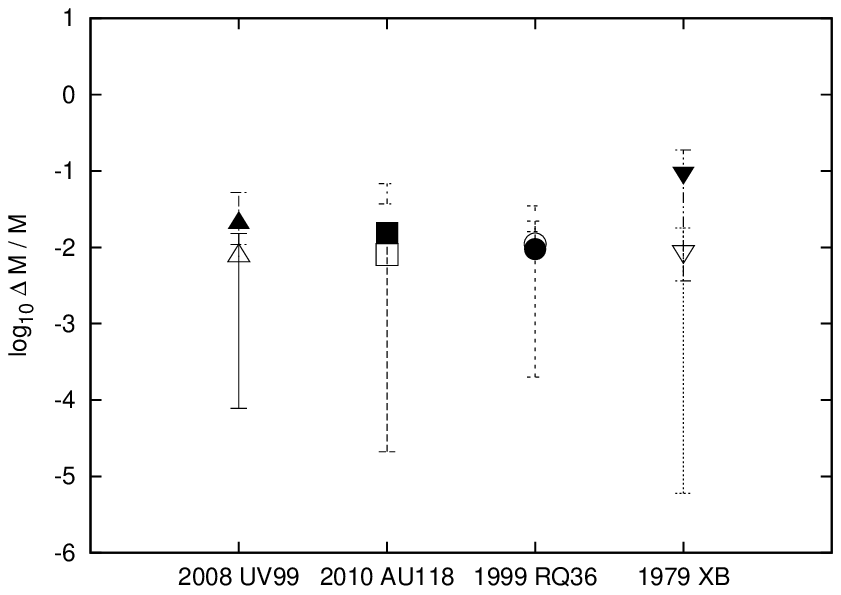} 
\end{tabular}\caption{Expected quality of orbital elements gained via two consecutive triangulations using simultaneous observations from Gaia and the Earth.
Medians and median deviations of the differences in actual and acquired osculating orbital elements are given for 4 NEAs. The open symbols denote the best case, and the full symbols the worst-case scenarios
given in Table~\ref{eggl:tab2}. The simulation time-span corresponded to Gaia's planned mission duration (5 years). \label{eggl:fig3}}
\end{figure}

In order to get a full set of orbital elements via TR, NEA velocities have to be estimated. As suggested in \cite{eggl-2011} a simple one-step interpolation approach is taken here, using two consecutively 
triangulated observations. This is possible despite Gaia's fixed scanning law, as there are two FOVs which are separated by an angle of $106.5^\circ$. Given an axial spin-rate of $60"s^{-1}$ the 
second FOV will pass over the same region as the first one with a delay of 106.5 minutes. However, Gaia's revolving scanning law includes additional 
precession as well as the drift caused by the satellite's proper motion around the Sun. Consequently, the two FOVs will only have an overlap between 0.01 and 0.04 square degrees - less than 10\% of the original FOV.
One could, therefore, argue that the prospects of observing an asteroid in both FOVs are rather small. 
To counter this argument DPSC's Gaia CU4 rendezvous simulator was used to determine whether the investigated asteroids  1999 RQ36, 2008 UV99, 2010 AU118 and 1979 XB will be observable by Gaia and cross both FOVs.
For 2008 UV99, 2010 AU118 and 1979 XB this was indeed the case for an initial phase angle of $0^\circ$. However, the actual crossing date as well the appearance in both FOVs depended on the scanning law's initial phase.
Given the TR method's strong dependency on the initial scanning phase for individual cases, a statistical argument may allow for a better evaluation of its applicability. 
\cite{bancelin-et-al-2010} conclude that independently of initial phase angles a total of  2180 NEOs and 585 Potentially Hazardous Asteroids (PHA) will be observed by Gaia. While precise
ranging data can be acquired for all of them using TR, two rapid consecutive observations would be required in order to generate a full set of orbital elements. 
Current estimates predict a fraction of approximately 20\% of Lead/Trail measurements for solar system objects (P. Tanga, 06/2013, private communication).
This population - roughly 400 NEOs - would allow for two consecutive TRs of positions, which in turn can be used to 
estimate the asteroids' velocity vectors. Without observational uncertainties the accuracy of the estimated velocities for NEOs would be on the order of $10^{-5} au/D$.
In order to see how the observational uncertainties in Table~\ref{eggl:tab2} together with the proposed velocity estimates influence an independent TR based determination of NEA orbital elements,
the simulation presented in the previous section was used to generate statistics on orbital element quality for the four selected NEAs (Table \ref{eggl:tab1}). 
Without knowing the initial phase of the Gaia satellite's scanning law, precise predictions on the quality of triangulated state vectors are not possible. 
Therefore, the median quality of triangulated osculating elements using two consecutive observations with $\Delta t=106.5$ simultaneously 
measured from Gaia and the Earth are presented in Figure \ref{eggl:fig3}. Only measurements in the method's accessible region (Figure \ref{eggl:fig1b}) during the
missions lifetime were considered, and then only when the target object was brighter than Gaia's limiting magnitude. 
Analogous to Figure \ref{eggl:fig4} best cases are depicted by empty, worst cases by full symbols (see also Table~\ref{eggl:tab2}).
Given a best case scenario, i.e. 0.1 arc-second precision Earth-based observations for the asteroids 2008 UV99, 2010 AU118 and 1979 XB, the expected quality of the orbital 
elements achievable is quite high.  
Especially in the case of 2008 UV99 the orbit uncertainties could be decreased by an order of magnitude compared to current values 
using two consecutive TRs only (cf. Table \ref{eggl:tab1}). 
If Earth-based observations are no better than arc-second precision such as assumed for the worst case scenarios, TR does not offer any substantial orbit refinement potential.
\begin{table}[b]
\begin{scriptsize}
\begin{tabular}{@{}lcccccccc@{}}
\textbf{Designation} & \textbf{H}  & \textbf{Nobs.} & \textbf{$\sigma_a$} & \textbf{$\sigma_e$} & \textbf{$\sigma_i$} & \textbf{$\sigma_\omega$} & \textbf{$\sigma_\Omega$} & \textbf{$\sigma_M$}\\
1999 RQ36  & 20.7 & 298 & 1.2$\cdot10^{-10}$& 3.4$\cdot10^{-8}$& 4.3$\cdot10^{-6}$& 6.4$\cdot10^{-6}$& 5.6$\cdot10^{-6}$&3.6$\cdot10^{-6}$ \\
1979 XB &  18.5   &  17   &0.24 &0.03 &0.82 & 0.37&0.06 &2.12\\ 
2010 AU118 & 17.9 & 19 & 0.57&0.04 &3 &50 &14 &89\\ 
2008 UV99    &    19.6 & 22 &0.30 &0.38 &27 &79 &18 &165 \\
\end{tabular}
\end{scriptsize}
\caption{NEA data from NASA JPL's Sentry risk table \cite{jpl-sentry-2012}. The uncertainties in the NEAs' orbital elements are shown in the corresponding $\sigma$ columns, 
where $\sigma_a$ is given in [au] and  $\sigma_{i,\omega,\Omega,M}$ are given in [deg]. $H$ corresponds to the absolute magnitude, and Nobs. are the number of observations.
\label{eggl:tab1}}
\end{table} 
\begin{table}[b]
\begin{tabular}{@{}ccccc@{}}
&\textbf{$\Delta \alpha_G$} & \textbf{$\Delta \delta_G$} &  \textbf{$\Delta \alpha_E=\Delta \delta_E$} & \textbf{$\Delta d_{EG}$}\\
best case & $3\cdot 10^{-4}$ & $1.8\cdot 10^{-3}$  & $0.1$ &  $150$ \\
worst case & $5\cdot 10^{-3}$ & $6\cdot 10^{-2}$    & $2$ &  $150$     \\ 
\end{tabular}
\caption{Dominating uncertainties for NEA orbit triangulation.
$\Delta \alpha_{G,E}$ and $\Delta \delta_{G,E}$ denote the respective angular uncertainties in arc-seconds [$"$] of Gaia \cite{hestroffer-et-al-2009} and Earth-based observations. 
$\Delta d_{EG}$ is the predicted in-mission uncertainty in the distance between Gaia and the Earth in [m].\label{eggl:tab2}}
\end{table}

\section{Statistical Ranging Constraints}
\label{eggl:sec:statr}
\begin{figure}[t]
\begin{tabular}{cc}
\includegraphics[scale=0.5]{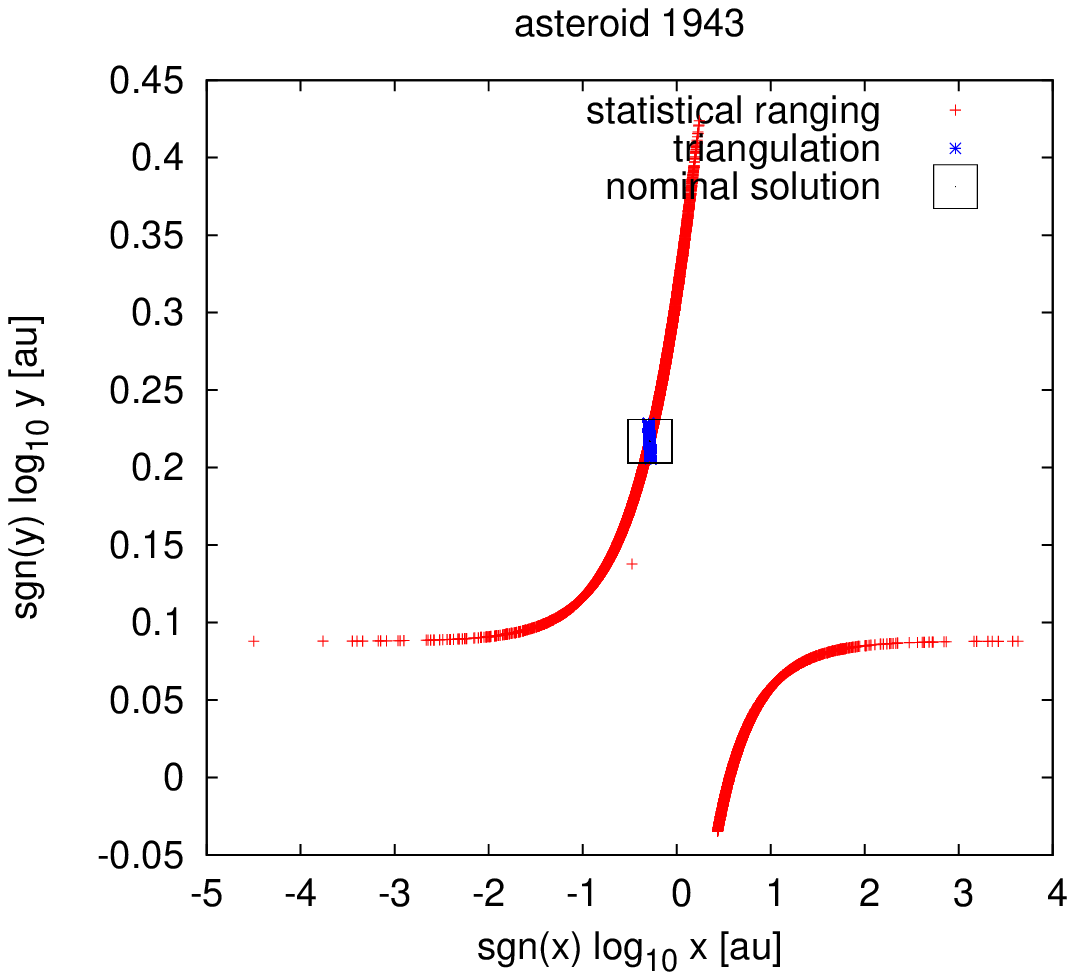} & \includegraphics[scale=0.5]{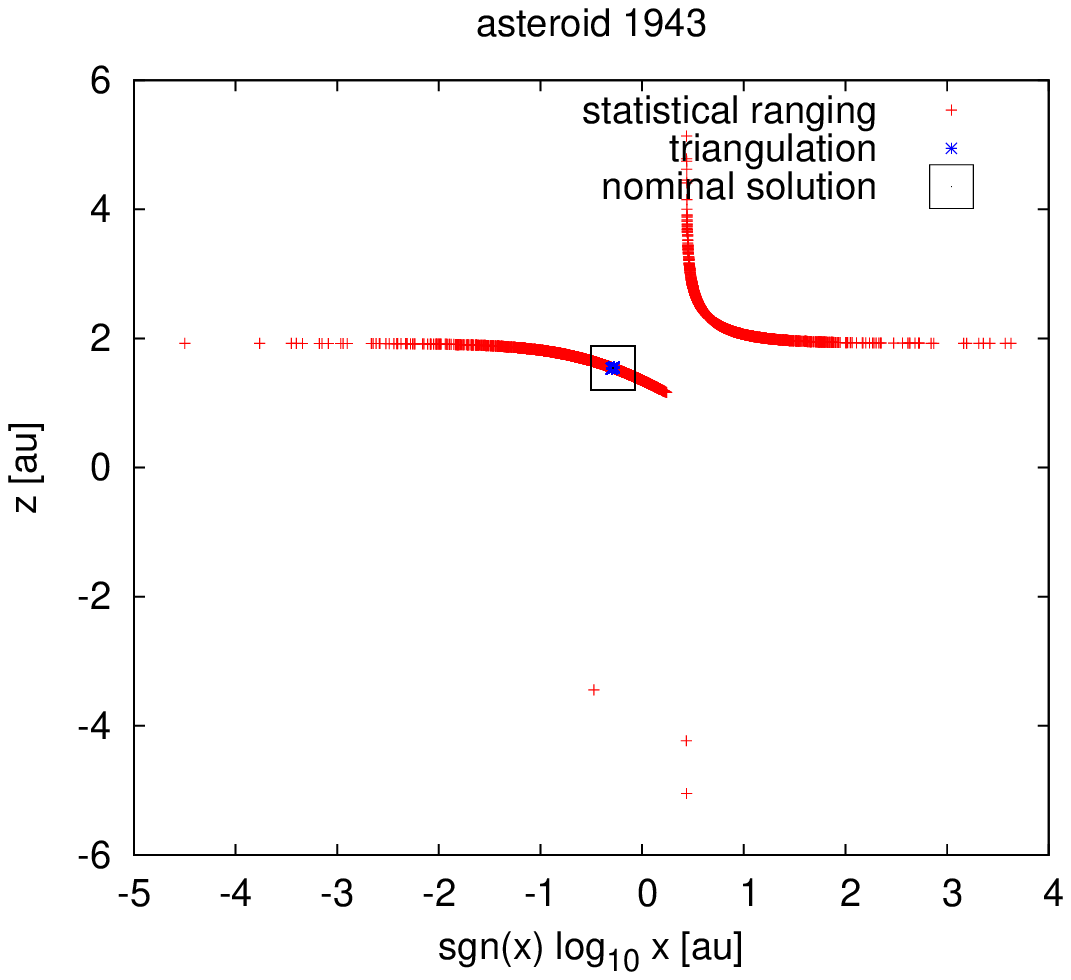} \\
\includegraphics[scale=0.5]{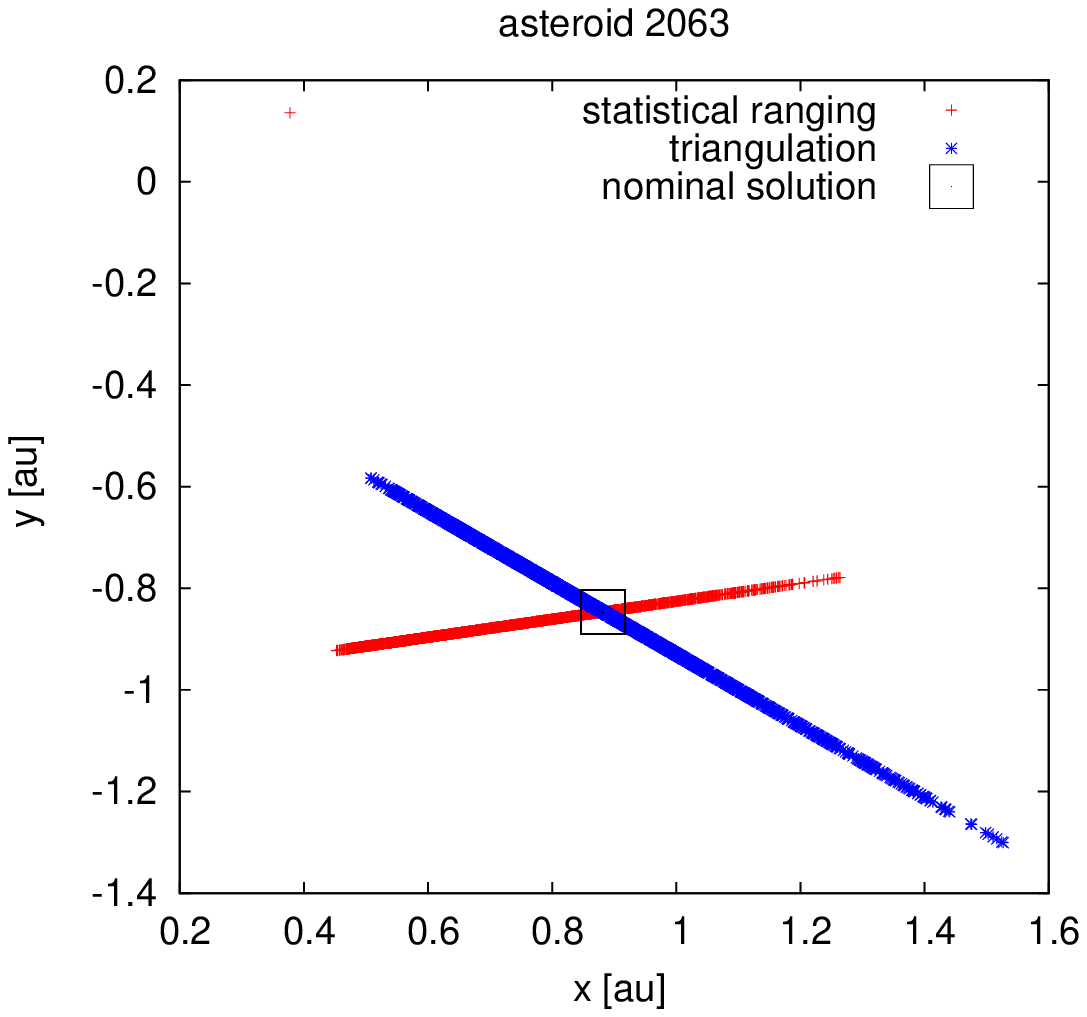} & \includegraphics[scale=0.5]{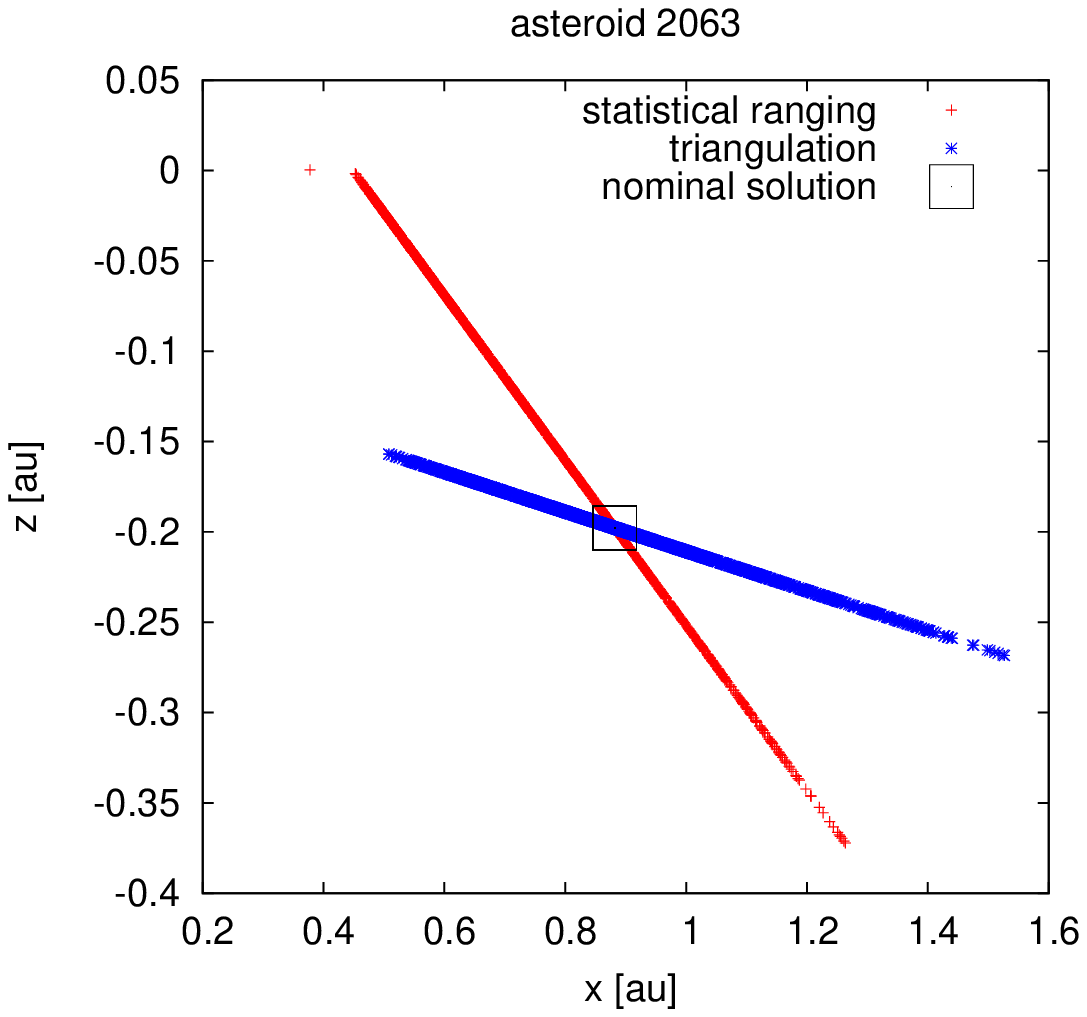} \\
\includegraphics[scale=0.5]{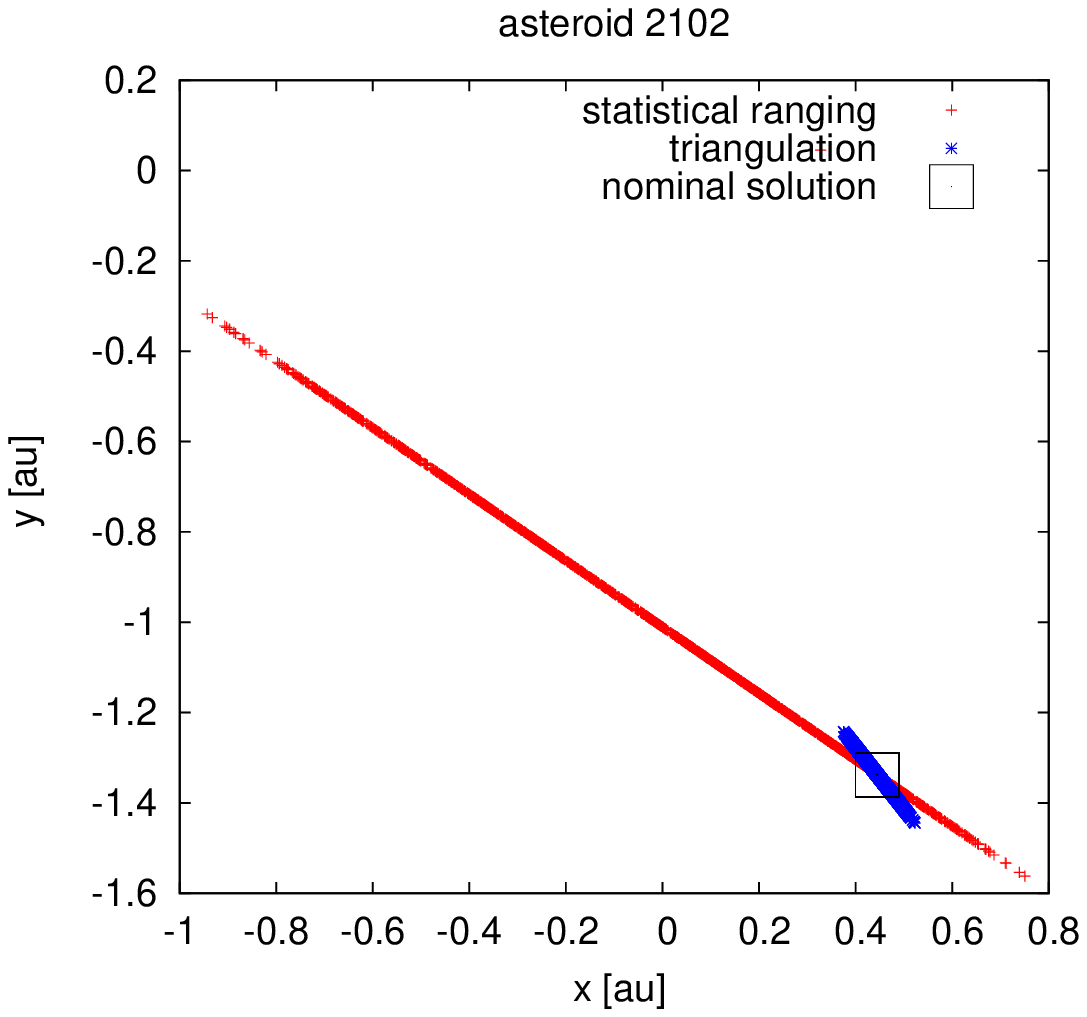} & \includegraphics[scale=0.5]{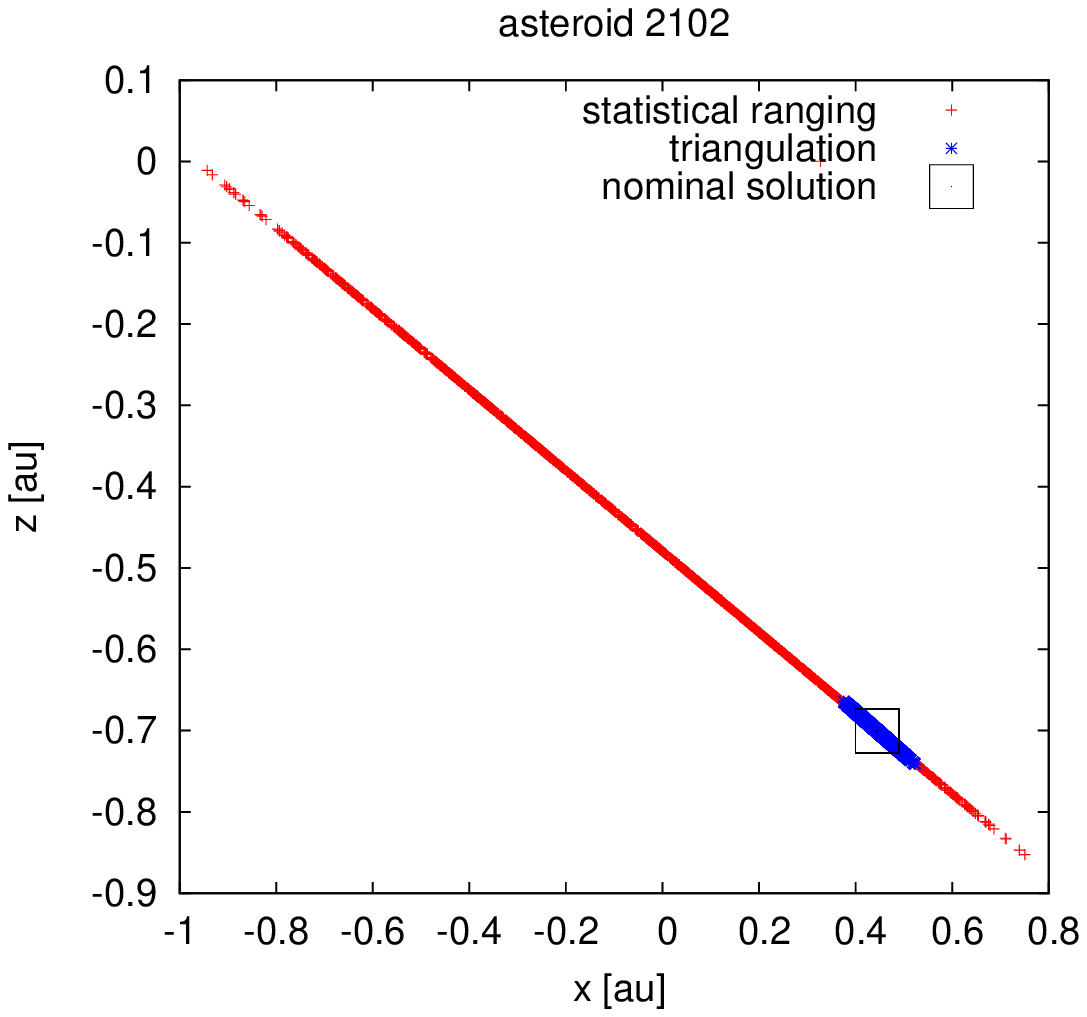} 
\end{tabular}\caption{Simulated localizations of three asteroids achieved via statistical ranging (SR) are compared to positioning via triangulation (TR).
The true orbit solutions are given by the black rectangle.
The possible range of TR results has been sampled from a uniform distribution, where worst-case scenarios have been assumed as defined in Table \ref{eggl:tab2}.
The scaling of the top panels is 'signed logarithmic', i.e. $-2$ corresponds to $-10^2=-100$ au. \label{eggl:fig4}}
\end{figure}

\begin{figure}[t]
\centering\includegraphics[scale=0.66]{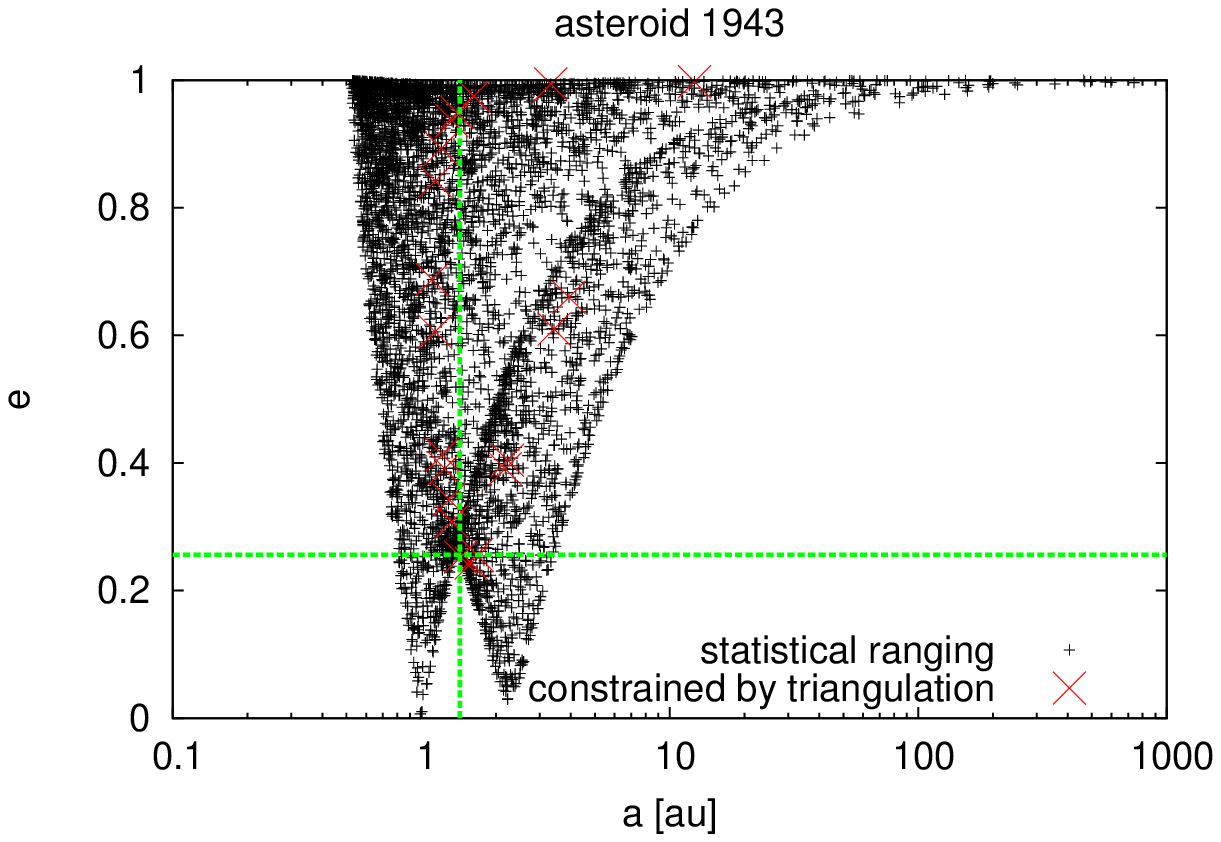} \\
\centering\includegraphics[scale=0.66]{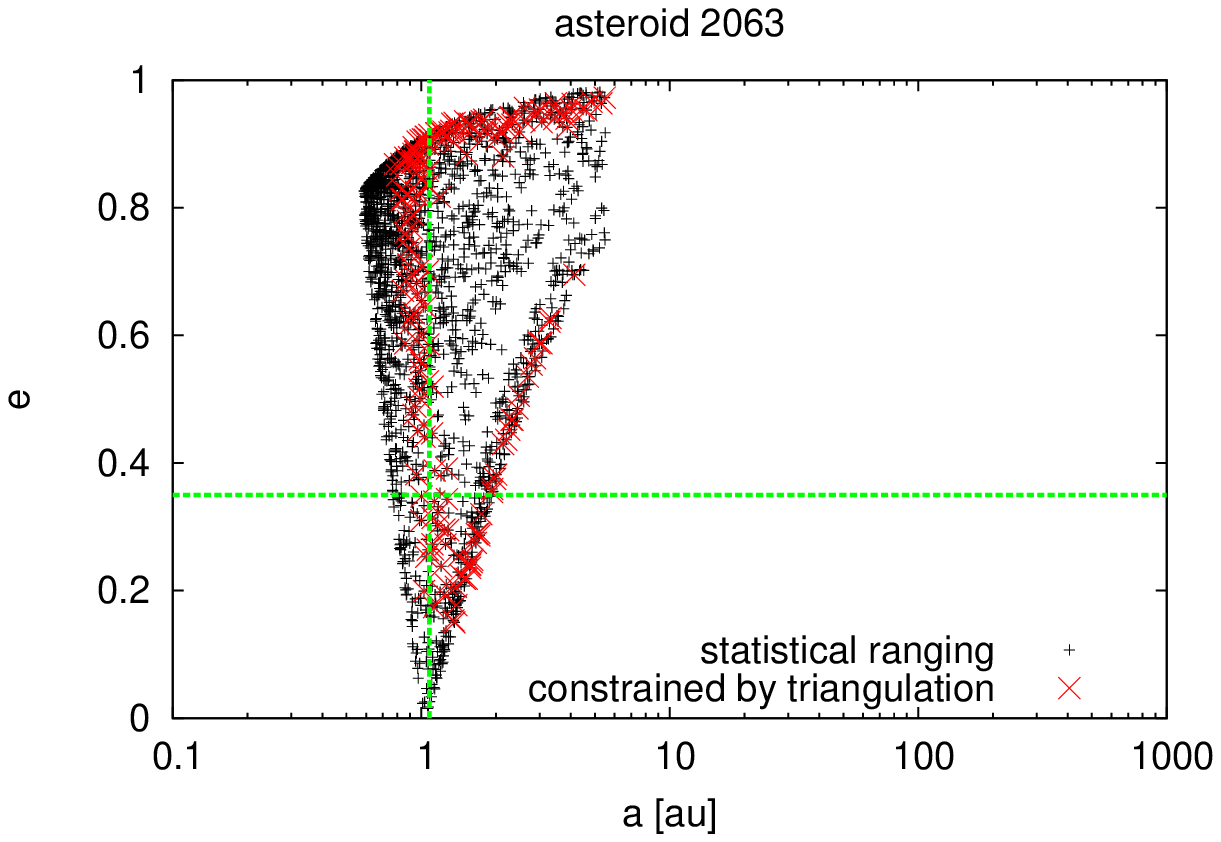}  \\
\centering\includegraphics[scale=0.66]{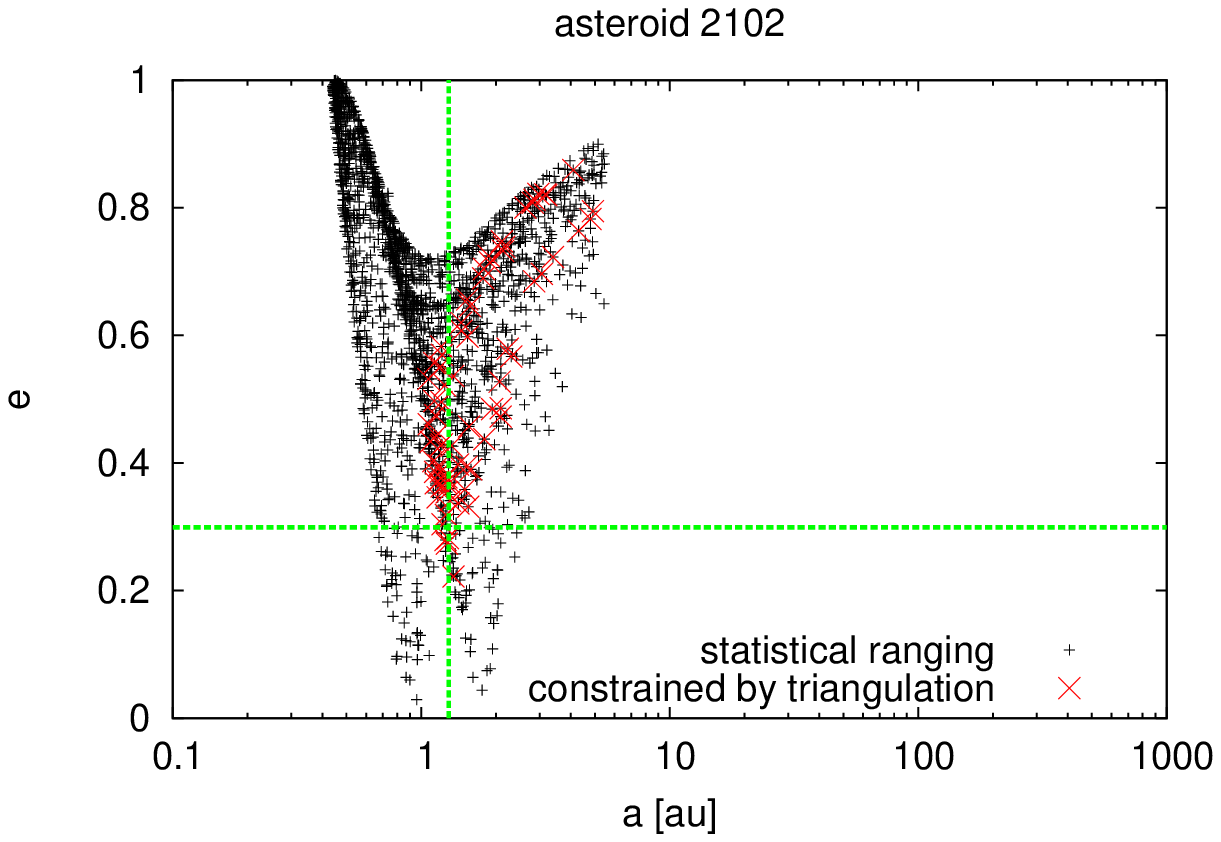} \caption{Possible semimajor axes (a) versus eccentricities of three asteroids are portrayed. Simulated statistical ranging (SR) results from Gaia data are shown as (+) signs.
The solutions that are constrained via triangulation (TR) from an Earth-based observer are denoted by ($\times$). The cross-hair gives the location of the true solution.
TR is capable of constraining SR orbital solutions considerably. \label{eggl:fig5}}
\end{figure}

The results of the previous section paint a rather bleak picture for TR regarding independent orbit determination and refinement.
Yet, the capability to produce observer-to-NEO distances provided by TR can be helpful in another way. 
Currently, the Gaia mission pipeline is intended to perform SR for newly discovered asteroids, if the observational data is not sufficiently
plentiful to allow for standard orbit determination.
Hereby, a set of possible orbit solutions compatible with the observed FOV positions are generated. The bundle of initial orbits can then be
propagated to generate ephemeris for follow up observations.
TR can be used to further constrain SR solutions by providing additional distance information.
We pointed out in section \ref{eggl:sec:pos} that observational errors prohibit an exact TR based localization of the asteroid. The uncertainties in the 
distance measurements would, however, be sufficiently small to constrain statistical ranging solutions.
This is portrayed in Figures \ref{eggl:fig4} and \ref{eggl:fig5}. Locations predicted by simulated SR and TR measurements are compared.
Figure \ref{eggl:fig4} shows the simulated positioning results from a single Gaia FOV crossing of the asteroids 1943 Anteros, 2063 Bacchus
and 2102 Tantalus. One can see that the intersection of the SR and TR based position sets contains the true orbit solution.  
Hence, those orbital solutions found by SR that are not compatible with TR solutions can be eliminated, see Figure \ref{eggl:fig5}. 
Even if the uncertainties in both methods are comparable, the combination of SR and TR can provide substantially more information on the location of the object.

\section{Discovery Cross-Matching}
Throughout the previous sections we have assumed that dedicated ground based sites are available to 
produce synchronized observations. This is certainly the most proliferating mode of 
operation, in particular regarding initial orbit determination. Yet, it is probably not
the most economic in terms of observational resources requirements. Therefore, one should consider alternatives,
such as using the results from large surveys. Potential discoveries of asteroids are
published very rapidly at the Minor Planet Center.\footnote{\url{http://www.minorplanetcenter.net/iau/NEO/ToConfirm.html}, 2013} 
We suggest that new discoveries or candidates observed from ground based facilities 
 be cross-linked with Gaia astrometric alerts. Should nearly simultaneous observations
be available, new objects can be linked via the methods discussed in section \ref{eggl:sec:idl}, 
and initial orbits may be constrained. 
As discussed previously, observations would not have to be perfectly synchronized.
Asynchronisities that cause the object to remain within the uncertainties caused by ground based astrometric precision
are permissible. For fast moving objects, however, this margin is small.
No accurate predictions can be given at this point 
on how often simultaneous observations of new discoveries will occur. Most likely they are rare events. 
However, discovery cross-matching comes at very little computational and no observational cost. Hence,
we consider cross matching a worthwhile effort.
A preliminary software procedure called GODSEND (Gaia and grOunD SurvEys for Neo Detections)
is currently implemented in the framework of the Gaia-FUN-SSO network.

\section{Conclusions} 
Simultaneous observations of NEOs from Gaia and ground based sites are valuable. 
They can be used to constrain statistical ranging results and facilitate identification and linking of observed objects.
Triangulation is a straight forward way to achieve such constraints, particularly when the number of available observations is small.
The TR method's potential for orbit refinement of known NEOs strongly depends on the
astrometric equipment available and the quality of results can vary from case to case.
Using the large parallax between Gaia and Earth-based observatories, substantial improvements of weakly constrained NEA orbits are possible, 
if a ground-based angular accuracy well below 1'' can be achieved.
In this context it might also be interesting to consider triangulation between Gaia and other contemporary space missions such as NEOSSat \cite{neossat-2012},
especially for orbital regions that are difficult to access from the ground.
Unfortunately, the actual NEO population available for a fully independent determination of orbital elements via triangulation alone may be as small as 400,
of which only 100 are PHAs. 
Therefore, a precise rendezvous predictions using the actual initial scanning law phase are necessary 
to draw a clearer picture on the possible merits of independent orbit generation via triangulation.
Nevertheless, synchronous observations from two sites have a substantial impact on the quality of preliminary orbital elements acquired via statistical ranging.
Even if no dedicated simultaneous observations were to be conducted from Earth based sites, Gaia discoveries can be cross-matched with the Minor Planet Center survey database 
in order to triangulate newly found objects at no additional observational cost.
\\\\
\textbf{Acknowledgments}
The authors would like to thank Daniel Hestroffer, David Bancelin, Paolo Tanga, Benoit Carry and Enrico Gerlach for their valuable input, as well as the DPSC for granting access to
the Gaia CU4 rendezvous simulator. 
Furthermore, the authors would like to acknowledge the support of the European Union Seventh Framework Program (FP7/2007-2013) under grant agreement no. 282703, as well as
the Gaia FUN-SSO network.

 \bibliographystyle{mn2e}
 \bibliography{eggl.bib}

%
%
%


\end{document}